# Designing Trustworthy User Interfaces for the Voluntary Carbon Market: A Randomized Online Experiment

Klaudia Guzij*
Center for Digital Technology and Management, Germany
klaudia.guzij@cdtm.de

Michael Fröhlich†
Center for Digital Technology and Management, Germany
froehlich@cdtm.de

Florian Fincke
Pina Technologies GmbH, Germany
florian.fincke@pina.earth

Albrecht Schmidt
Ludwig Maximilian University, Germany
albrecht.schmidt@ifi.lmu.de

Florian Alt
Bundeswehr University Munich, Germany
florian.alt@unibw.de

## ABSTRACT

The voluntary carbon market is an important building block in the fight against climate change. However, it is not trivial for consumers to verify whether carbon offset projects deliver what they promise. While technical solutions for measuring their impact are emerging, there is a lack of understanding of how to translate carbon offset data into trustworthy interface designs. With interaction between users and offset projects mainly happening online, it is critical to meet this design challenge. To this end, we designed and evaluated interfaces with varying trust cues for carbon offset projects in a randomized online experiment (n=244). Our results show that content design, particularly financial and forest-related quantitative data presented at the right detail level, increases the perceived trustworthiness, while images have no significant effect. We contribute the first specific guidance for interface designers for carbon offsets and discuss implications for interaction design.

## 1 INTRODUCTION

Carbon offsetting is becoming increasingly popular for reducing greenhouse-gas emissions and limiting the climate catastrophe [40, 56]. The voluntary carbon market – individuals or companies offsetting emissions without regulations requiring them to – can become an important building block in the fight against climate change [5]. For individuals to engage with carbon offset projects it is important that they understand and trust the projects they invest in. With carbon offset projects spread across the globe and the long time period over which the impact takes effect, it is not trivial to verify whether projects deliver what they promise. With public demands for more transparency [10], there is considerable ongoing effort to develop data-driven systems that allow the tracking of projects in a transparent an verifiable way (e.g. [37]). However, one aspect neglected so far is how to present the vast variety of possible information pieces about carbon offset projects so users can evaluate their trustworthiness. This is particularly relevant for interaction design research as online interfaces are the primary way of how information on carbon offset projects is communicated. As offset buyers cannot check the status of the projects themselves the establishment of trust needs to be mediated through user interfaces. The increasing demand for transparency and upcoming new ways to measure data on carbon offset projects result in a need to explore how to design trustworthy interfaces for them.

HCI research has identified ease of use, content design, and visual design as important aspects to influence trustworthiness of user interfaces [50, 64]. With no literature addressing interfaces of carbon offset projects, the field of charitable giving offers a starting point. Similar to offsetting, donors give money voluntarily and it is difficult to verify the impact of any particular donation. Charities can improve trust by increase transparency regarding both financial [91] and non-financial data [6, 73]. Recent work shows that interface design can help establish a trust relationship between donors and charities [50]. In particular, content design has a strong influence on the perceived trustworthiness of charitable actions in online contexts [50, 82]. As content design is domain-specific, insights cannot be transferred to carbon offsetting directly. Thus, while we can assume a similar relevance of content design to build trustworthy interfaces, we do not yet know which data can serve as reliable trust cues for user interfaces of carbon offset projects.

To address this gap, we set out to explore interface content design at the example of forest-based carbon offset projects. We designed website interfaces combining findings from HCI research, a qualitative pre-study (n=13), and sampling of existing websites of established carbon offset sellers (n=7). With these interfaces, we tested the effect of different types of data (financial, images, forest) and their level of detail (low, high) on perceived trustworthiness in a randomized interactive online experiment (n=244). Participants of the experiment were asked to evaluate the trustworthiness of four forest carbon offset projects and select one project to invest in. The interfaces for each project differed concerning the type of data and the level of detail they provided. Our results show that not all data types are equally effective as trust cues. Quantitative forest data – such as number of planted trees and captured carbon – stands out as the strongest predictor of perceived trustworthiness and project selection. Our results also show the importance of assessing the level of detail for each data type separately. For example, at a high level of detail the interface showing finance-related data was chosen significantly more often and received significantly higher ratings of trustworthiness than at a low level of detail. Providing images of the tree planting progress did not affect project choice or perceived trustworthiness of the project.

Our results are in line with research from the field of charitable giving, where quantitative data was shown to improve perceived trustworthiness of charitable projects. From our results we derive and discuss theoretical and practical implications for researchers and interface designers. Doing so, this paper is the first to report specific guidance for interface designers on what data may serve as trust cues in the domain of carbon offsets. More generally, our findings shed new light on the question of how designers can create online experiences that facilitate trust through interfaces in situations where users cannot check the fulfillment of a transaction directly. The growing interest in the voluntary carbon market along with the public demand for more transparency will increase the need to improve interfaces of carbon offset projects. Our research points to the importance of focusing HCI research on trust-building user interfaces that guide users through complex data with appropriate design. Future work may build on the findings to develop guidelines on how to transparently communicate the progress of carbon offset projects.

**Contribution Statement:**

The contribution of this work is threefold. (1) It explores different types of trust cues for carbon offset projects and presents six website interfaces. (2) Using these interfaces it evaluates the influences of these trust cues, specifically the type of data (financial, images, forest) and their level of detail (low, high), in a randomized online experiment (n=244). (3) Finding significant differences related to project choice and perceived trustworthiness, it discusses theoretical and practical implications and is the first work to report design guidance related to building trustworthy interfaces in the domain of carbon offsets.

## 2 BACKGROUND & RELATED WORK

Our work draws from several strands of research, most notably from the fields of charitable giving and trustful interfaces.

### 2.1 The Voluntary Forest Carbon Market

Carbon offsetting refers to the process of funding the avoidance of emissions through climate protection projects and thus compensating for carbon emissions that arise from industrial or other human activities [10, 44]. Carbon offset projects are increasingly demanded in the so-called voluntary carbon market, where private organizations, communities, and individuals buy credits with the aim to reduce their carbon footprint to claim „carbon neutrality" without regulation demanding them to. Offset sellers can either be private or governmental organizations that manage administrative processes of a climate protection project. [10, 68]. While there are several types of carbon offset projects this paper focuses on forest-based ones. Forest-based projects account for the highest share in nature-based carbon offset projects, in which offsetting is performed through avoided deforestation (emission avoidance) or afforestation (emission sequestration) [66]. Despite their potential and impact, academic literature and press coverage have documented various challenges and critical problems of forest carbon offsets, especially in the context of project selection and data communication [54, 55, 78]. Particularly, a lack of transparent data has been a topic of current reports as offset buyers cannot easily access data in order to verify a project's effectiveness [10, 23, 37].

While there are ongoing efforts to meet these challenges and establish more transparent [] and verifiable [] methods to measure the impact of carbon offset projects, we were not able to identify research looking at the human side – i.e. what information needs to be provided by interfaces so people can evaluate the trustworthiness of carbon offset projects. With a public demand for more transparent communication, this makes forest-based projects an interesting example at which explore the design of interfaces that facilitate the establishment of trust.

### 2.2 Trustful User Interfaces

*Trust* is a broadly used concept. In the following we provide a definition for the context of this work and summarize existing research on trustworthy user interfaces.

*2.2.1 Definition of Trust and Trustworthiness.* Trust is a multi-faceted concept, combining emotional, cognitive, and behavioral dimensions [12, 89]. Cho et al. studied the multi-faceted meanings of trust and define it as follows [13] :

*"Trust is the willingness of the trustor (evaluator) to take risk based on a subjective belief that a trustee (evaluatee) will exhibit reliable behavior to maximize the trustor's interest under uncertainty (e.g. ambiguity due to conflicting evidence and/or ignorance caused by complete lack of evidence) of a given situation based on the cognitive assessment of past experience with the trustee." (p. 28:5)*

Sheppard & Sherman further identify key aspects of why and when trust is needed and under which circumstances it is formed, based on various multidisciplinary contexts [77]. Generally, trust is created from vulnerability that is needed in the presence of risk and interdependence. [51, 70, 77]. Trustors (the ones trusting) will only accept vulnerability if they have strong expectations that a certain outcome will be achieved and that the trustee (the one trusted in) does not exploit their vulnerability [2, 30]. Sociological research, as well as research in computation and automation, state that the need for trust is based on the goal to reduce complexity and uncertainty [41, 53].

According to various researchers, trustees have to fulfill specific trustworthiness beliefs for trust to be established. Being antecedents to trust, the three most relevant trustworthiness beliefs are *competence*, *benevolence*, and *integrity* [31, 32, 57, 62, 71, 74, 84, 85]. Competence describes the ability to perform an action the trustor needs; benevolence means that the trustee acts in the trustor's best interest; integrity deals with the trustee's honesty and promise-keeping [58]. This paper adopts a multidimensional definition of trust and thus sees the three beliefs as antecedents that need to bet met in order to establish trust.

*2.2.2 Trustworthy User Interfaces.* Trust and trustworthy interfaces have been a topic of interest in interaction design research and at DIS (see e.g. [17, 18, 42, 90]). Online contexts require trustors to rely on their trustee's computer-mediated communication (their website), which ultimately increases the complexity of trust [1, 9, 43, 75]. Whereas the previously mentioned antecedents of trust represent cognitive factors that can be applied more generally, the context of online interfaces adds a layer of functional factors that influence trustworthiness. The main factors that influence trustworthiness in online contexts are *ease of use*, *content design*, and *visual design* [84]. *Ease of use* examines the effort that a user feels when using an (online) system [19, 20]. Trustors who find the navigation of a website easy perceive it as more trustworthy [82]. However, research also found that there is a subjective factor to perceived ease of use [38]. *Visual design* refers to the aesthetic appeal of a website [41]. *Content design* focuses on creating trustworthiness by providing information that is perceived as useful and comprehensive [76]. Trustworthiness of user interfaces increases when they provide both comprehensive information and convey expertise and honesty, without creating bias [25, 53].

Scholars studied the interplay of these three antecedents of trustworthiness [64, 82, 83]. Pengnate and Sarathy drew from Norman's emotional design framework and found that visual appeal is important to create initial trust signals for a website's first impression [61, 64]. Thielsch et al. go beyond first impressions and study determinants making users revisit and recommend a website. For complex decisions involving higher cognitive processes the importance of content design increases, whereas aesthetics have a lower level of importance [82, 83]. Stanford et al. studied the difference between consumers and experts. Consumers rely on website aesthetics when assessing trustworthiness of websites, unlike experts, who rather focus on factors related to content design [79].

For content design it is important to note that it is application- and domain-specific which kind of information is perceived as *useful and comprehensive*. Approaches to generalize findings about content design across domains have resulted in ambiguous findings, which can be seen at the example of images as content type. On the one hand, scholars focusing on social presence cues state that trustworthiness can be increased by using images as they create a feeling of face-to-face interactions [25, 70, 81]. On the other hand, other studies could not find an effect of images on trustworthiness of user interfaces [69, 70].

Drawing from these findings we consider content design to be the most relevant factor for building trustworthy interfaces for carbon offset projects. Given the domain-specific nature of content design, this in turn highlights the importance of exploring what constitutes useful and comprehensive information in the realm of voluntary carbon offsetting.

*2.2.3 Measuring Trustworthiness of User Interfaces.* There are several approaches measuring trustworthiness in online contexts [34, 71]. We build on a measurement instrument developed by McKnight et al. that addresses trust as a multi-dimensional concept specifically in the context of websites [58]. Their instrument is based on four trust-related constructs, namely the institution-based trust, the disposition to trust, trusting intentions, and trusting beliefs [58]. For the context of this work, we focus on the construct of trusting beliefs (competence, benevolence, and integrity). Various HCI researchers used and adapted the measurement instrument by McKnight et al. to measure perceived trustworthiness in online contexts [11, 13, 74]. Focusing on website design, Secker et al. found that high integrity and competence lead to higher trustworthiness, and an untrustworthy experience is related to missing benevolence and integrity [74]. Hence, benevolence and integrity are necessary yet not sufficient to establish trustworthiness on their own. For websites to be perceived as trustworthy competence is need as well. These findings underline that perceived trustworthiness should be measured by including all three constructs, as the combination of benevolence, integrity, and competence has proven to be the most reliable way to derive a holistic trust measurement.

It is important to note that the perceived trustworthiness of an interface does not say anything about whether the carbon offset project represented in the interface is actually deserving of trust. We acknowledge that by exploring how to build trustworthy interfaces in the context of carbon offsets, there is a risk that the results can be used to apply manipulative design techniques. Manipulative design and dark patterns are, unfortunately, a reality and have been a subject of interest at DIS (see e.g. [7]). However, we can only detect, stop and prevent manipulation if we know what elements can be used to manipulate. We believe that publishing open and accessible research on trustworthy interfaces is thus ultimately of public interest.

## 2.3 Charitable Giving

Charitable giving refers to any act of providing money, time, or goods with no value or monetary return for the donor [88]. Donors face a dilemma as they have to trust a charitable organization while not being fully able to monitor their performance and quality [67]. The actions of charities and outcomes of donations are perceived as highly intangible [65]. This phenomenon can also be observed with carbon offsets, as they also face a missing tangible counter value at which a user can verify that a transaction succeeded. Interest within the HCI community in the field has picked up (e.g. [4, 22, 86, 87]) as donations are increasingly performed and reported online and donors increasingly demand more transparent communication.

*2.3.1 Trust in Charitable Giving.* Trust research in the field of charitable giving focuses on contextualizing and recognizing trust in the relationship between individuals and charitable organizations [36, 92]. Research found that traditional factors such as brand and reputation, familiarity, and geographic distance strongly influence donations [24, 35, 73]. However, studies show that there is a shift in requirements on what makes charities trustworthy. With media reports on charity scandals claiming donations not being used for their cause, the public increasingly demands greater accountability and transparency [24]. Donors face a dilemma as they have to trust a charitable organization while not being fully able to monitor their performance and quality [67]. Donor questionnaires increasingly state that charitable organizations should go beyond the (legal) required minimum of reporting to increase public trust [91]. Yang and Northcott derived that the public asks explicitly for *direct* accountability through continuous communication to donors [91]. Sargeant and Lee also validated this measure [73]. Next to financial transparency, non-financial information has been identified as critical to improve trustworthiness [6, 73]. Non-financial data can be information on the *outputs* and *outcomes* that a charity achieved. While *outputs* are defined as the direct result of the input, e.g. number of fed children, *outcomes* are defined as the indirect benefits or changes in, e.g. condition, status, or skills of the beneficiaries [16].

*2.3.2 Trustworthiness of Charitable Giving in Online Contexts.* As charitable giving is increasingly performed online, charities are faced with the challenge of establishing trust cues in their online communication. As online contexts allow anyone to publish any information, traditional credibility cues such as the reputation are not sufficient to assess a charitable organization's trustworthiness. The increasing demand for transparency and online communication have thus started to motivate the HCI community to analyze online representation of charitable projects. Yang and Northcott analyzed how charities communicate their information and identified missing comprehensiveness of communicated data [91]. Krasteva and Yildirim examined the process of getting informed before and after a donation [49]. They found that the main reason why only a few people research before donating lies in the effort to retrieve and process information. They highlight the importance of access to data as there is an increasing number of donors for whom it is not sufficient to donate based on the feeling of "warm glow" – a better emotional state of mind after donating. With these "rational donors" in mind, the authors recommend charities to improve their website communication [49].

In the context of website design, scholars studied trustworthiness of charitable organizations based on the elements of visual design, ease of use, and content design of their websites. Robins et al. found that visual design judgments play a highly significant role in rating charities' credibility [72]. Küchler et al. examined the effect of website design perceptions on willingness to donate [50]. They created an extension of the technology acceptance model by van der Heijden and investigated the influence on a user's intention to donate [20, 50]. Their findings state that content design has the highest impact on willingness to donate. These results are in line with findings of Thielsch et al. arguing that content design increases in importance when donating one's own money compared to someone else's, as higher cognitive processes are being activated [83]. These findings explain why choosing the right type of content becomes increasingly important for charitable organizations to improve trustworthiness of their projects. Based on these findings, we assume an equally important role of content design to create trustworthy interfaces for carbon offset projects.

## 2.4 Summary

Voluntary carbon offset projects have the potential to contribute to the fight against climate change. As most communication on the progress and impact of offset projects happens online, this poses the challenge of designing user interfaces that facilitate trust. Drawing on HCI literature we hypothesize that trust can be facilitated through online user interfaces if they are perceived as trustworthy (competence, benevolence, and integrity). Specific to online interfaces we know that visual design, content design and ease of use influence users' perceived trustworthiness of websites. With many parallels to voluntary carbon offsetting, research about website design of charitable organizations offers a starting point to explore the design of trustworthy carbon offset interfaces. Both fields address users that give voluntarily and the impact of transactions is realized in distant locations or over long time periods. This makes it difficult to verify whether trust was not misused. Research in the field of charitable giving identified content design – providing information that is perceived both useful and comprehensive – as the most influential factor for establishing trustworthiness and influencing the willingness to donate. Based on these findings, we assume an equally important role of content design to create trustworthy interfaces for carbon offset projects. As content design is highly domain specific, the main goal of this paper is to explore which types of content can serve as reliable trust cues and how to present them in user interfaces of carbon offset projects.

## 3 RESEARCH APPROACH

To determine potential trust cues for carbon offset interfaces, we first conducted semi-structured interviews to examine the buyer-seller trust relationship and understand requirements of both sides. We then sampled websites of existing carbon offset sellers to understand the existing design space. While the results of these steps are not at the core of this paper, we think it is worth reporting how they contributed to the establishment of our hypotheses.

### 3.1 Interview Study

We conducted semi-structured interviews with offset buyers (n=7) and sellers (n=4) to examine the buyer-seller trust relationship and understand requirements for communicating progress of carbon offset projects, with specific focus on forest carbon offsets. Our goal was to build an initial qualitative understanding of the space, its stakeholders, and their relationships. Grounded Theory provided a systematic approach for simultaneous data collection and analysis with a low risk of including researchers' bias [33]. The small sample size naturally limits the generalizability of these findings. From the interview data we identified three main challenges related to the trust relationship between buyers and sellers, namely (1) choosing a trusted seller for buyers, (2) providing transparent data on the progress of carbon offset projects, and (3) communicating updates in an easily understandable and accessible way. We identified four areas in which data can be used to communicate the progress of offset project in a more transparent way:

- **Geographic focus**: Information on the exact location and size of the carbon offset projects, visualization of the project area, e.g. through satellite images.
- **Financial transparency**: Information on costs involved in carbon offset projects, including a split of costs and the different stakeholders getting paid through the donation.
- **Forest-related measurements**: Continuous and easy-to-understand updates on additionality (a comparison to a baseline scenario in which the investment had not happened), accumulated biomass, biodiversity measurements, and a clear measure of how much carbon is stored in the project.
- **Images**: Images (or videos) of updates about the project. A particular interest was set on images of the forest area, tree planting activities portraying workers, as well as evidence of the tree growing progress.

### 3.2 Design Space Analysis

To better understand the existing design space for forest carbon offset user interfaces, we additionally sampled websites of (n=7) forest carbon offset sellers (see supplementary material). There are a variety of design approaches for communicating project progress, using different data types and levels of detail. User interfaces of established organizations (e.g. ClimatePartner, South Pole) tend to provide mainly qualitative data with little detail. Interfaces included text-based descriptions, annual carbon offset capacity, the area's contribution to the Sustainable Development Goals (SDGs), and images or videos providing context around the project. Images typically portrayed different stages of forest growth, different types of plants or habitats of animals, and sometimes plantation activities, with a focus on portraying the inclusion of local communities. Only few providers focused on more detailed approach, using quantitative data to communicate project progress over time. Those who did were startups (e.g. Ecologi, Pachama) rather than established incumbents and they focused either on forest-related data or financial data. Interestingly, quantitative data was not always integrated into the user interface of the specific project website. For example, Ecologi redirected to a hosted spreadsheet to provide access to the raw financial data set. Figure 1 shows an overview of design approaches.

### 3.3 Conceptual Framework

The results of our interview study and design space analysis indicate a demand for and a trend towards interfaces that use quantitative data to provide transparent updates on projects. While interfaces of incumbents rely largely on qualitative data (e.g. images) to inform about projects, emerging companies are trying to integrate quantitative data (e.g. financial data and forest data) to build trust through transparent updates. This trend is interesting, because research in online donations found that content design – i.e. what and how to present content in interfaces – to be particularly important for cognitive actions that go beyond the first impression [50]. Building these insights, we identified (1) the type of data and (2) the level of detail at which to present it to be interesting dimensions to be investigated as trust-building factors.

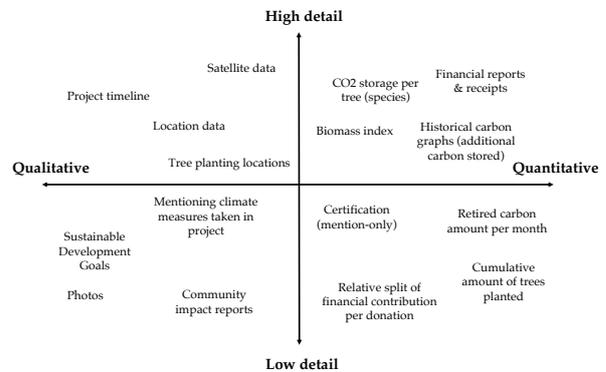

Figure 1: The design space for online interfaces in the voluntary forest carbon market. While more recent companies experiment with quantitative approaches, incumbents mainly rely on qualitative data.

*Type of Data.* We theorize that the type of data portrayed on a user interface impacts the selection of carbon offset projects and their evaluation of trust. We operationalize this variable through three instances: financial data, images, and forest data. Financial data transparency emerged throughout the interview study, design space analysis, and literature from charitable giving [6, 73, 91]. Likewise, images were both mentioned by interviewees and are widely used in existing interfaces. Prior HCI research, however, is inconclusive on their effect on trust, making them an interesting target to investigate [69, 70, 81]. Finally, quantitative forest data emerged in both the interview process and the design space analysis.

*Level of Detail.* Our analysis of existing interfaces shows that there are few examples that visualize quantitative data at a high level of detail. If data is provided at high level of detail it was not well-integrated into the interface. The interview study indicates that buyers wish for more detailed information, specifically on quantitative data. At the same time, sellers in our interview study, aiming to provide more transparency, assumed limited interest in quantitative data from buyers. These contradictory assumptions raise the question of whether the level of detail affects project choice and perceived trustworthiness. Thus, it is interesting to evaluate how to integrate highly detailed quantitative data into the user interface. As content design focuses on creating trust by providing information that is perceived as useful and comprehensive, we see the level of detail as a decisive dimension in designing user interfaces for carbon offsets.

### 3.4 Hypotheses

We theorize that the type of data (financial, image, forest) and the level of detail at which it is presented (low, high) in user interfaces influence the trustworthiness users perceive towards a forest carbon offset project. As dependent variables we investigate both project choice by participants in a fictitious scenario as well as the perceived trustworthiness. A detailed description of the dependent variables can be found in the Method section.

Focusing on the type of data as independent variables we define the following hypotheses:

**Hypothesis 1:** *The type of data impacts the choice of a forest carbon offset project.*
**Hypothesis 2:** *The type of data impacts the evaluation of trust in a forest carbon offset project.*

Additionally, we suggest that the level of detail of the portrayed data will impact project choice and perceived trustworthiness. Therefore, we posit the following hypotheses:

**Hypothesis 3:** *The level of detail impacts the choice of a forest carbon offset project.*
**Hypothesis 4:** *The level of detail impacts the evaluation of trust in a forest carbon offset project.*

## 4 METHOD

We conducted a randomized online experiment with an interactive prototype to test the effects of the (1) type of data (financial, image, forest) and (2) level of detail (low, high) on the project selection and perceived trust. The following section describes our method.

### 4.1 Participants

We recruited 244 participants via university mailing lists, reaching approximately 7000 recipients and sharing it on social media. The entire experiment was conducted in English, and participants were briefed about an approximate completion time of 10 minutes. To increase participation, we committed to donating 0.20 EUR per completed survey to an offsetting project and raffled three vouchers worth 20 EUR among participants. Our final sample consisted of 244 participants, out of which 54.1% were male, and 44.3% were female. The average age of our participants was 27.41 years. 95.1% of our participants lived in *anonymized*. Our participants were mainly students (59.8%), followed by employees (32.4%), self-employed people (6.1%), and retirees (1.6%).

### 4.2 Apparatus

*4.2.1 Independent Variables.* Based on the definition of our independent variables and our hypotheses, we designed seven different user interfaces using Figma[1] and deployed them as interactive website with Useberry[2]. Each interface described a forest carbon offset project with text-based project information (title, location, project description). Six out of seven interfaces showed additional information about the project by providing either financial, image, or forest data at low or high levels of detail. We were careful not to include any other difference in the interface or project description that could influence the experiment. Figure 2 shows screenshots of the six interfaces. The control project and overview page are not depicted, but can be found in the supplementary material.

---

[1] https://www.figma.com/ (last accessed 2021-08-09)
[2] https://www.useberry.com/ (last accessed 2021-08-09)

*Interface 1A/ 1B: Financial Data.* Financial data provides additional information on how each Euro invested in the project is spent. We communicated that 40% of the donations are needed for administrative overhead, and 60% are invested into reforestation activities and visualized the information as an annotated donut chart. In the low level of detail interface (1A), this was the only information provided. The high level of detail interface (1B) portrays a more granular split with direct financial indication to the different cost blocks of the 40/60 split. The relative cost block indication was derived from research on cost splits of the clean development mechanism [68].

*Interface 2A/2B: Images.* We analyzed images of existing providers to identify commonly used image content. We portray the progress of tree planting by showing images of plants in different types of progress and included plantation workers in the images. We did not include images that might trigger strong emotional responses, such as direct-facing portraits, families, or children. In the low level of detail interface (2A), one image is shown. The high level of detail interface (2B) shows additional images to provide more context.

*Interface 3A/3B: Forest Data.* Forest data refers to quantitative information on the development of the specific forest within the project. We focused specifically on the number of trees planted and the annual tons of carbon offset. In the low level of detail interface (3A), the total of planted trees and hectares of forest protected and the annual carbon offset are shown. The high level of detail interface (3B) shows additional information about the type of trees planted, a biodiversity index, and a comparison graph modeling the annual amount of offset carbon with and without the project from its beginning to 2060.

*Interface 4: Control.* To create a standard comparison basis for the experiment, we took inspiration from the descriptions of the sampled carbon offset sellers and created a control project. All elements of the control project are also present in the other interfaces. We altered only the structure and formulation of sentences so that participants did not feel reading the same text across all projects. We were careful to describe the project in a similar language and format as found for real projects. The interface describes the project with a title, project id, location name, and a tag on a map. A text-based description explains the project setup, providing a qualitative report summary as typically done at established carbon offset sellers. It consists of a problem statement and a description of the project's activities, with specific mentions that the project benefits biodiversity or local community education. To avoid bias through geographical proximity we located all projects in the Amazon Rainforest region of South America, expecting our participant sample to come predominantly from *anonymized* [35].

*4.2.2 Dependent Variables.* The two dependent variables were the selected project and perceived trust in each project.

*Project Selection:* After reviewing all interfaces, each participant had to select one project in which to invest. We measured the project selection as a dichotomous variable for each project [0=No;1=Yes].

*Perceived Trust:* After selecting one project, participants were asked to rate each project on a five-point Likert scale. Following the example of previous research on trust in online contexts [11, 13, 74] we adapted three questions from the Trusting Beliefs instrument by McKnight et al. [58]. To acknowledge the multi-dimensionality of trust, the averaged scores of the three questions represented the final trust value index.

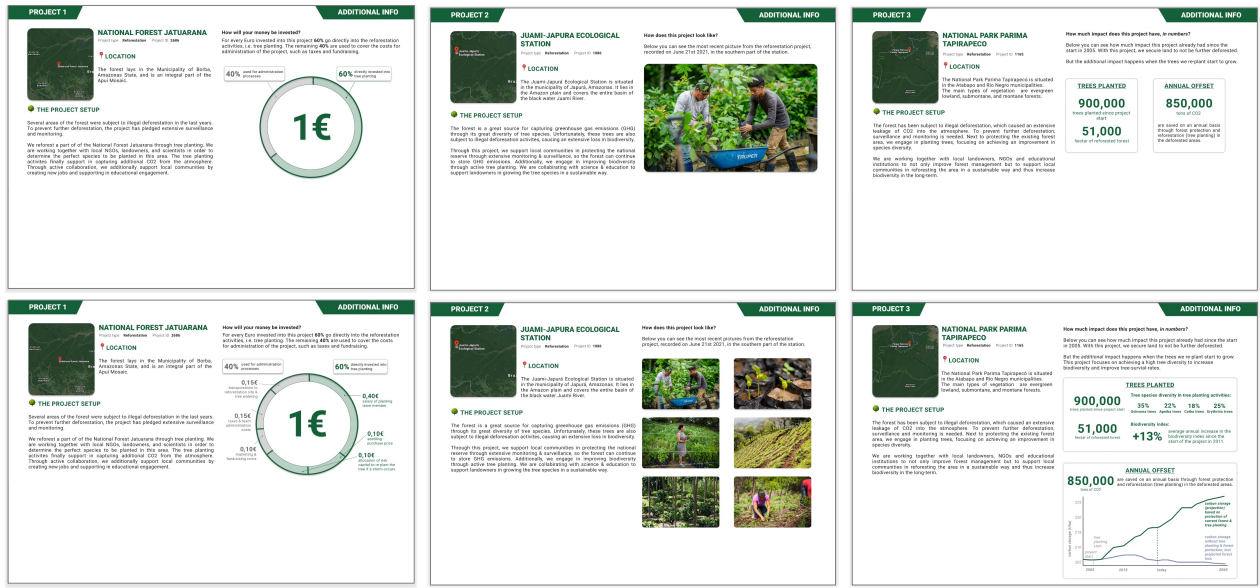

Figure 2: The prototyped interfaces. Each interface shows a general text-based project description on the left side and additional data on the right side. The additional data operationalizes the independent variables, namely the type of data and the level of detail at which it is presented. The first column from the left shows financial data, the second one images, and the third one forest data. The first line shows the low level of detail and the second line shows the high level of detail. High resolution images are in the supplementary material.

(1) *I believe this project would act in my best interest.*
(2) *I would characterize this project as honest and trustworthy.*
(3) *This project performs its role of communicating updates on their progress very well.*

### 4.3 Experimental Design

The design of our experiment uses a split-plot setup, utilizing both between-group and within-group components [52]. To investigate the influence of the data type, we adopted a within-group design – each participants was shown interfaces with with each type of data (financial, image, forest data, control). To investigate the influence of the level of detail, we adopted between-group design – each participant was shown all interfaces either from the low-detail (Group A) or from the high-detail (Group B) condition. Table 1 shows the different experimental conditions. As both groups also contain a control project, the total number of conditions in this experiment is seven. Each participant was shown exactly four interfaces (one for each data type). To avoid sequence and carryover effect for the within-group design, we adopted a Latin square. The order of the conditions was systematically arranged in a counterbalanced order, and participants were randomly assigned to the conditions [52]. We were careful to avoid any difference between the interfaces aside from the independent variables investigated (see supplementary material for screenshots of all interfaces).

Table 1: Overview of the experimental conditions. Participants saw the control project and either interfaces from Group A or B.

|  |  | within group | | | |
|---|---|---|---|---|---|
|  |  | Financial | Images | Forest | Control |
| between | low detail | 1A | 2A | 3A | 4 |
| group | high detail | 1B | 2B | 3B | |

### 4.4 Procedure

The experiment was administered online with the survey tool Useberry[3], allowing interactive website-like interaction with the prototyped interfaces. The data was collected anonymously. The experiment only enabled participation for desktop devices. Participants in total had to complete six steps: (1) answer a brief entry survey, (2) read the scenario description, (3) browse through four projects, (4) select one project, (5) evaluate the trust of each project, and (6) answer a demographic survey. The scenario provided to participants stated that they should imagine being customers of a bank committed to becoming climate-neutral. As customers, they are asked to select in which project the bank should invest on their behalf. Given recent media reports, we asked participants to pay attention about choosing a trustworthy project.

## 5 RESULTS

In total 244 participants completed the experiment. Given the split-plot design, our sample can be divided into two groups: Group A, which saw the low detail versions of the projects (n=119), and Group B, which saw the high detail versions of the projects (n=125). The sample in both groups mainly consists of students (58% & 61.6%) with a similar average age (28, SD=8.228 & 27.5, SD=7.139). The average completion time of the experiment was 9.25 minutes.

---
[3]https://www.useberry.com/ (last accessed 2021-08-09)

### 5.1 Behavior of Overall Sample

Participants' project choices show differences between the low detail and high detail sample. Table 2 shows an overview of the selected projects by group. Both groups decided on Project 3 (forest) with the highest proportion (55.5%, 43.6%). Project 1 (financial) scored the second. While participants in the low detail group selected Project 1 24.4% of the time, the proportion almost doubles to 40% in the high detail group. Project 2 (image) ranks third in the high detail group (9.6%). In the low detail group, however, it was selected only 8.2% of the time – less often than Project 4 (control). Project 4 (control) was selected 11.9% in the low detail group and 4.8% of the time in the high detail group.

**Table 2: Overview of the project selection. The relative frequencies at which projects were selected by Group A (low detail) and Group B (high detail)**

|  | Group A (low detail) | Group B (high detail) |
|---|---|---|
| Project 1 (Financial) | 24.4% | 40.0% |
| Project 2 (Images) | 8.2% | 9.6% |
| Project 3 (Forest) | 55.5% | 43.6% |
| Project 4 (Control) | 11.9% | 4.8% |

After selecting a project, we asked participants to evaluate the trustworthiness of each project. Figure 3 provides an overview of participants' trust index ratings. The results show that Project 3 (forest), which was also selected most often, scored the highest across both groups (Mode 3.3 & 3.7; Mean 3.7 & 3.4). Project 1 (financial) is perceived more trustworthy in the high detail group. For Project 2 (image) there is a strong deviation between mode and mean in Group A (2.7 & 1.66), resulting in a left-skewed distribution that indicates discrepancy in opinions regarding the trustworthiness. Project 2 (image) showed no differences between groups, while Project 4 (control) is evaluated with lower trust in Group B. Here, we also see a left-skewed distribution (2.3 & 1.66).

### 5.2 Hypothesis Testing

The collected data consists of different types of variables. Since all of them are ordinal measures, they do not meet the assumptions of parametric tests. Therefore, we tested our hypotheses with non-parametric tests [52].

*5.2.1 Type of data & choice of project:* H1 deals with the question of whether the type of data [financial, image, forest] impacts the choice of forest carbon offset projects. We conducted the Cochran's Q test [14]. It compares more than two dependent variables based on the ranks of the dependent variable for significant differences to verify whether $k$ treatments have identical effects. The data comparison was made within groups 1 and 2 separately.

*Group A (low detail).* There is a statistically significant difference between the types of data, $X^2(3)=65.639$, p<.001. As the Cochran Q test only states the existence of a difference, a post hoc test determined the exact difference between the projects. There is a statistically significant difference between Project 2A [images] and 3A [forest] (p<.001); Project 4 [control] and 3A [forest] (p<.001); as well as Project 1A [financial] and 3A [forest] (p<.001). Project 2A [images] and 4 [control] (p<.604); Project 2A [images] and 1A [financial] (p<.014); likewise Project 4 [control] and 1A [financial] (p<.052) showed no statistically significant difference. Our pairwise comparison shows that Project 3A [forest] was the leading indicator for the difference in Group A.

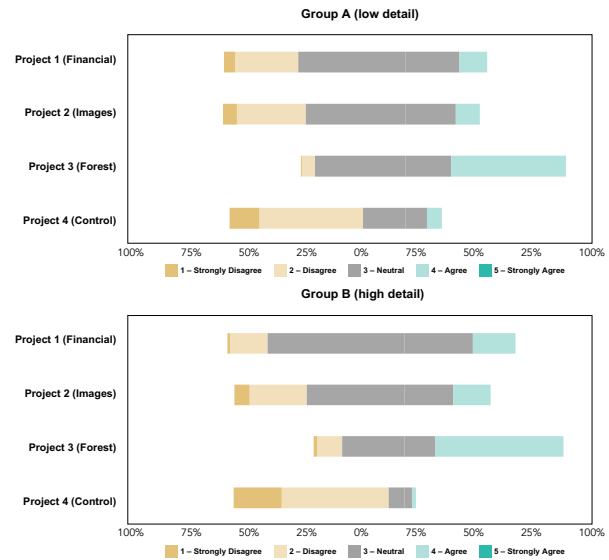

Figure 3: The trust index for each project by group. Participants rated the trustworthiness of each project by answering three questions on a five-point Likert scale. A trust index was calculated as the mean value of the three answers. The diverging stack bar charts provide an overview of participants' responses by group and project. Comma values were rounded to generate the graph.

*Group B (high detail).* There is a statistically significant difference between types of data, $X^2(3)=64.729$, p<.001. The post hoc test shows different results in the pairwise comparison of group 2. There is a statistically significant difference between Project 4 [control] and 1B [financial] (p<.001); Project 4 [control] and 3B [forest] (p<.001); Project 2B [images] and 1B [financial] (p<.001); as well as Project 2B [images] and 3B [forest] (p<.001). Project 4 [control] and 2B [images] (p<.448); likewise Project 1B [financial] and 3B [forest] (p<.376) showed no statistically significant difference. In summary, we can state the following:

- Group A: Reject $H1_0$, Retain $H1_1$
  – *The type of data [financial, image, forest] impacts the choice of a forest carbon offset project.*
- Group B: Reject $H1_0$, Retain $H1_1$
  – *The type of data [financial, image, forest] impacts the choice of a forest carbon offset project.*

*5.2.2 Type of data & evaluation of trust:* H2 deals with the question of whether the type of data [financial, image, forest] impacts the evaluation of trust in forest carbon offset projects. This hypothesis was initially tested within-group, using Friedman's two-way analysis of variance, a non-parametric alternative to the one-way repeated measures ANOVA [52]. It checks for a difference in multiple dependent samples and the population from which these have been drawn [52]. If the Friedman test is statistically significant, we can say that there is at least one difference between one of the variables. Our results show a statistically significant difference between the control project, financial data, forest data, and images when evaluating the trust of a carbon offset project, $\chi^2(3)$ = 325.471, p <.0001. The post hoc analysis shows that the pairwise comparison is statistically significant between all projects, except between Project 2 [images] and Project 1 [financial], which are not statistically significant (p <0.12).

As these results differ from the project choice analysis, we conducted a second round of Friedman tests separating Group A and Group B. The separate analysis shows that both groups would reject the null hypothesis (p < .001). However, the post hoc pairwise comparisons show differences between the groups. Group B, the high detail group, is different across all pairwise comparisons with statistical significance. Hence, in the high detail group the type of data makes a difference when evaluating the trust of a forest carbon offset project. Group A, the low detail group, shows the same results as the cross-group analysis. All pairwise comparisons show statistically significant differences, except for Project 2 [images] and Project 1 [financial], which achieve no statistical significance in their pairwise comparison (p < .422). The result of this in between-group differences will be addressed in H4. In summary, we can state the following:

- Reject $H2_0$, Retain $H2_1$
  - *The type of data impacts the evaluation of trust in a forest carbon offset project.*

### 5.2.3 Level of detail & choice of project.
H3 deals with the question of whether there is a difference in low level and high level of detail when users choose their preferred offset project. We performed a between-group analysis to analyze frequency counts between Group A and Group B. We used the chi-square test of homogeneity $\chi^2$ [80]. There was a statistically significant difference in the multinomial probability distributions between the two groups (p = .024). Therefore, we can reject the null hypothesis and accept the alternative hypothesis. Our cross-tabulation results show that people in Group B choose Project 1 [financial] (13.7% vs. 63.3%) more often than in Group A. People in Group B also choose Project 3 [forest] less often than in Group A (53.7% vs. 46.3%). Project 2 [images] was selected more often in Group B (45.5% vs. 54.5%). Project 4, our control group, was selected significantly less often in Group B (70% vs. 30%). In summary, we can state the following:

- Reject $H3_0$, Retain $H3_1$
  - *The level of detail impacts the choice of a forest carbon offset project.*

### 5.2.4 Level of detail & evaluation of trust:
H4 assumes in its $H_0$ that the level of detail does not impact the evaluation of trust in a forest carbon offset project. Our dependent variable is the trust index, calculated as mean of the three trust questions answered on an ordinal five point Likert scale. Each project is analyzed separately.

The level of detail (1 for low, 2 for high) represents our dichotomous independent variable. Statistical analysis does not directly test for association between a dichotomous independent variable and an ordinal dependent variable [8, 80]. However, research suggests running an ordinal logistic regression and using the coefficient (of the dichotomous variable) from this analysis to establish a statistically significant relationship between these two variables using an odds ratio [52, 80]. The odds of an event occurring is the probability of it occurring versus the probability of it not occurring.

*Project 1 (Financial).* Before looking at the result of the odds ratio for project 1, it is necessary to check the overall model fit, i.e., whether the data sample is appropriate for the regression model [80]. The Pearson as well as the deviance goodness-of-fit test indicated that the model was a good fit to the observed data, $\chi^2(11)$ = 10.573, p < .480, $\chi^2(11)$ = 11.537, p < .399. The likelihood-ratio test, also called the model-fitting information, evaluates the change in model fit by comparing the full model to the intercept-only model. The final model statistically significantly predicted the dependent variable over and above the intercept-only model, $\chi^2(1)$ = 10.634, p < .001. This means that the independent variable adds to the prediction of the dependent variable. The odds of seeing low level of detail and rate a high trust value was 0.476 (95% CI, 0.304 to 0.745), a statistically significant effect, $\chi^2(1)$ = 10.537, p < .001. Since the regression weight (B score) shows a negative value (-0.742), this means that people who have a low detailed prototype are not likely to evaluate Project 1 with a high trust score. The higher the regression weight, the likelihood of rating a high trust value increases.

*Project 2 (Images).* The Pearson as well as the deviance goodness-of-fit test indicated that the model was a good fit to the observed data, $\chi^2(11)$ = 12.185, p < .350, $\chi^2(11)$ = 13.323, p < .273. The likelihood-ratio test did not achieve the desired statistical significance, $\chi^2(1)$ = 162, p < .688. This result means that the independent variable does not account for the prediction of the dependent variable. The odds of seeing low level of detail and rate a high trust value was 0.914 (95% CI, 0.590 to 1.416), a not statistically significant effect, $\chi^2(1)$ = 0.162, p < .687. Although the regression weight shows a negative value (-.0.90), we do not have significant results to state that a low detailed prototype can predict a high trust score for Project 2. Thus, it does not make a difference if it is a low detailed or high detailed prototype for the trust evaluation of Project 2.

*Project 3 (Forest).* The Pearson as well as the deviance goodness-of-fit test indicated that the model was a good fit to the observed data, $\chi^2(9)$ = 14.086, p < .119, $\chi^2(9)$ = 15.059, p < .089. The likelihood-ratio test did not achieve the desired statistical significance $\chi^22(1)$ = 0.162, p < .813. This result means that the independent variable does not account for the prediction of the dependent variable. The odds of seeing low level of detail and rate a high trust value was 1.055 (95% CI, 0.678 to 1,640), a not statistically significant effect, $\chi^2(1)$ = 0.56, p < .813. Even if the regression weight shows a positive value (.053), we do not have significant results to state that a low detailed prototype can predict a high trust score for Project 3. Thus, it does not make a difference for the trust evaluation of Project 3 if it is a low detailed or high detailed prototype.

As two projects showed a difference in level of detail when evaluating trust, we can summarize the following:

- Reject $H4_0$, Retain $H4_1$
  - *The level of detail impacts the evaluation of trust in a forest carbon offset project.*

## 6 DISCUSSION

This study set out to explore how to design trustworthy interfaces in the context of the voluntary carbon offset market. Our experimental results show that the choice and the evaluation of trustworthiness in a forest carbon offset project depends on the type of data portrayed in the user interface and its level of detail. We found that not all data types are equally effective as trust cues. Additionally, the right level of detail must be assessed for each data type separately, as a higher level of detail does not automatically increase trustworthiness for each data type.

Our study results are in line with research from the field of charitable giving, where content design, with specific focus on quantitative data shows to have the strongest influence on perceived trustworthiness of charitable projects. This study is the first to report specific guidance for interface designers on what data may serve as trust cues in the domain of carbon offsets. More generally, our findings shed new light on the question of how designers can create online experiences that facilitate trust through interfaces in situations where users cannot check the fulfillment of a transaction directly. In the following we discuss both theoretical an practical implications for interaction design.

### 6.1 Overview

We identified quantitative forest data as the strongest predictor for evaluating a forest carbon offset project as trustworthy and ultimately choosing it. Our pairwise comparisons with all other data types showed that participants selected projects reporting forest data significantly more often. A low level of detail, namely showing the protected area and the annual offset volume was already sufficient to establish trust into the project. Projects that provided financial transparency were chosen the second most often and ranked second in the evaluation of trustworthiness. The level of detail at which data was made available made a significant difference for interfaces showing financial data. By providing more detail, the proportion of people selecting the project almost doubled from 24.4% in Group A to 40% in Group B. This result, for one, highlights the positive impact of financial transparency on trustworthiness and confirms findings from the field of charitable giving [6, 73, 91]. However, it also sheds light on the fact that establishing trust through user interfaces depends on not only the type of data, but also how it is presented. Projects providing images portraying the forest area or tree planting activities did not affect trustworthiness or project choice. This is in line with literature, finding that images can create a broad range of reactions, from suspicion to enthusiasm [69]. Our data indicates that the images we used did not affect trust in the context of forest carbon offsets. However, images with different content – i.e., emotional content, progress of a specific tree, pictures contextualized with specific captions – might lead to different results. Given contradictory findings from literature, it could be interesting to revisit the impact of images and other qualitative trust cues in combination with quantitative data, as they could improve trustworthiness as a complementary data source.

The most relevant question in the context of this paper probably is the following: Does more transparency and providing more data in user interfaces lead to increased trust in forest carbon offset projects? In summary, our results show that we indeed can facilitate trust through online interfaces. Our results also point to the importance of contextualizing content design. It is crucial to understand which data is essential within the relevant domain context and how to present it. In the following, we discuss and present both theoretical and practical implications for HCI and interaction design that arise from these findings.

### 6.2 Theoretical Implications

From our results we can draw several theoretical implications.

*The increasing demand for voluntary carbon credits and lack of transparency in today's market will lead to increased pressure to provide more transparent and continuous data on carbon offset projects.* The voluntary carbon offset market is facing accelerated growth in demand. While established companies rely on their brand reputation, new market entrants will need to find ways to compete with them for consumers' trust. From our interview study and design space analysis, we can speculate that new offset companies will use more transparent and continuous data updates to establish trust and by doing so create the brand value to challenge incumbents. At the same time – with a similar development already today in charitable giving – we expect consumers to push for more transparency of carbon offset projects. Given the missing standardization, we suggest that the HCI community should pursue an active role in further understanding and establishing guidelines on presenting and integrating data into user interfaces. Along with technical innovations enabling better tracking of projects' impact, it is equally important to develop design standards to present the data so that it can be easily understood and interpreted by users. Particularly, the field of Sustainable HCI has the potential to go beyond interactions between humans and computers and include interdisciplinary research to define design approaches based on psychological and economic decision models [3, 15, 46].

*Integrating additional data into user interfaces of carbon offset projects facilitates the establishment of trust in online contexts.* Our results show that providing additional data on carbon offset projects supports the trustworthiness of these projects. Our results connect to literature on charitable giving that highlights the importance of data transparency [91]. However, not all information is equally suited to facilitate trust. For interfaces of forest carbon offset projects, quantitative data on a project's development, such as forest or financial data, outperforms qualitative data, such as images. This advantage of quantitative data is likely to transfer to other domains and projects, such as other types of carbon offsetting, plastic offsetting, as it already created a shift in the field of charitable giving in general. Given the domain-specific nature of trust cues, it is necessary to identify what type of data is appropriate for each context.

*How data is presented in online interfaces impacts trustworthiness.* Our results show that it is not enough to only present the right type of data. It is also important to do it in the right way. Different types of data are affected differently by this relationship. For financial data, communication at a too superficial level does not result in increased trustworthiness. Instead, it might rather raise users' suspicions if the provided data appears not to be comprehensive. For forest data, on the other hand, interface elements providing low-detail information already had a positive effect. In such a case, presenting data in a too detailed format may well decrease the desired effect as it overloads users with data they are not familiar with interpreting. Consequently, it is necessary to analyze each type of data independently to identify the appropriate level at which to present it. While our research focused on the level of detail to investigate, it is likely that there are more dimensions (i.e. visual design, ease of use) that are relevant.

### 6.3 Practical Implications

From our findings we derive three implications for practitioners and interface designers.

*Experiment with continuous quantitative data in user interfaces:* Our study results show that quantitative data supports the establishment of trust better than qualitative data. Literature from charitable giving further highlights the importance of continuous updates to increase donor lifetime value [21]. From this we can infer that users will develop more trust, if they receive updates about the impact of their carbon offsets over time. Quantitative data is more difficult to interpret for users, thus practitioners will need to meet this design challenge through experimentation with different data types, their combinations, and visualization. The results of our study suggest that focusing on inputs, outputs, outcomes [26] related to financial and forest data may be practical starting points to communicate continuous progress. Providing data in machine-readable formats would allow integration into other products and services, which, in turn, would help create transparency. We recognize that developing interfaces with frequently updated information comes with many implementation challenges but want to highlight it as an opportunity to build long-term brand value in a growing market.

*Build interfaces that guide users' sense-making process to establish trust:* Our research points to the importance of domain experts, i.e. offset sellers, to closely collaborate with user interface designers to find approaches to visualize data appropriately. We highlight the role of user interface designers in guiding users' sense-making process when being confronted with quantitative data on carbon offset projects. Given the complexity of carbon offsets, it is important to find the appropriate abstraction level to present different data types. We suggest starting at a high abstraction layer that communicates the most important metrics and allows users to understand the big picture, but then allowing them to drill down to see more detailed data if interested. At the example of financial data interfaces this can be realized in charts portraying high-level financial information of a project, allowing users to deep dive into specific financial indicators by interacting with the chart. Ultimately, such an approach would support serving different user groups and allow to work around trade-offs such as between intuitive understandability and comprehensiveness [45]. Finally, these findings also underline the importance of creating interactive interfaces that allow users to access their desired data points from one single source, without overloading them.

*Design for different user groups: warm-glow and rational offsetters:* From literature in charitable giving, we know that donors can be distinguished into "warm-glow" donors and rational donors [47]. Similarly, people are likely motivated by different reasons to offset. From our pre-study and the ambiguous results of image data interfaces, we speculate that there are both "warm-glow" offsetters and rational offsetters. Whereas rational offsetters are interested in understanding the impact of their investment, warm-glow offsetters are motivated by the "feeling of a warm-glow" after investing. It might be worth considering these target groups and their needs independently when designing user interfaces. Drawing from research in the field of persuasive technologies for sustainable behavior change, we believe that targeted design has additional potential to create change behaviors and attitudes of users towards a more rational and transparency-focused assessment of data [59, 60].

### 6.4 Future Research Directions

Like any empirical study, ours is not without limitations. Our sample is not representative of the general *anonymized* population. More importantly, we simulated the selection of the carbon offset project, and, therefore, no real investment by participants had to be made. In the following, we outline how future work may overcome these limitations and expand on our work.

- **Field Study:** We propose to run a study in collaboration with an established carbon offset seller, aiming for a representative sample. For one, this would allow us to investigate consumers' motivation to offset and evaluate our hypothesis in a real-world setting. It would be interesting to further segment participants to understand whether different groups of offsetters exist. For example, it would be interesting to understand whether prior knowledge about offset markets impacts the perceived trustworthiness of interfaces.
- **Further Trust Cues:** Our experiment is limited to three data types and two levels of details. Future work might expand on both dimensions to better understand trust cues for carbon offset projects. For example, certification labels and geographical location might be good starting points.
- **Interaction Effects:** We investigated the effect of each data type in isolation. Future work could address the interaction between different data types on perceived trust and project selection. Specifically, the role of image data as a complementary data source could be an interesting starting point.
- **Dynamic Data Exploration:** Drawing from the state of current interfaces for carbon offset projects, we conducted the experiment with static data presented in the interfaces. Interaction design research could investigate the impact of interactive interface elements that allow users to explore data themselves.
- **Verifiable Data:** While this work looks into what data types and what presentation facilitate trust, we do not address how users might verify the correctness of the presented data [45]. The collaboration between many stakeholders makes this a sizable challenge. Future work might address this from both a user-centric and a technical perspective. New technologies such as blockchain might be well suited to provide transparency on a technical level [39, 48, 63]. However, blockchain systems themselves remain difficult to use for both beginners [27, 28] and established users [29]. HCI research should explore how to build interfaces that remove these barriers and allow users to verify their offsets.

- **Manipulative Design and Dark Patterns:** With initial insights on how content can be adapted to create trustworthy interfaces for carbon offsets, we acknowledge the associated risk of players with false motivation being enabled to create manipulative designs and promote carbon offsets that do not deliver upon their promise. The application of HCI methods can be used to understand the factors of content design that influence users to become victims of dark patterns and establish metrics on how to work against them.

## 7 CONCLUSION

This paper examines which data users need to evaluate the trustworthiness of forest carbon projects, and investigates how this data can be best translated into online user interfaces. Our work was motivated by the growing voluntary carbon market and consumer demand for more transparent communication. The results of our randomized experiment reveal that by providing the right data in the right way user interfaces can facilitate the establishment of trustworthiness. They also show that not all data types are equally effective as trust cues and the level of detail at which data is presented in the interface is relevant depending on the data type. In our experiment, integrating quantitative forest-related and detailed financial data in the user interface increased trustworthiness, whereas images on their own had no statistically significant effect. Rooted in these findings, we present both theoretical and practical implications for researchers and designers. Doing so, this study is the first to report specific guidance for interface designers on what data may serve as trust cues in the domain of carbon offsets. On a more general level, our research points to the importance of user interfaces that guide users through complex data with appropriate design, especially in cases where there is a high uncertainty if a certain outcome has been achieved or not. We derive from our findings that contextual research in content design will become increasingly important for HCI research in the field of non-tangible services. With this work we also hope to increase efforts enabling offset buyers to better deal with information delivery on carbon offsets and identify manipulative designs. Whereas knowing about domain-specific content design artifacts creates a starting point to include additional mechanisms to create trust through transparency in the field of carbon offsets.

## ACKNOWLEDGMENTS

This work was supported by the Deutsche Forschungsgemeinschaft (DFG) (grant no. 316457582 and 425869382). We thank the team from https://condens.io/ for supporting us with their qualitative research analysis tool — it helped us analyze and understand the interview data we collected.